\documentclass[prd,superscriptaddress,nofootinbib,twocolumn]{revtex4-2}
\usepackage[T1]{fontenc}
\usepackage[utf8]{inputenc}
\usepackage[english]{babel}
\usepackage{microtype,amsfonts,amssymb,amstext,amsmath,mathtools,physics,xcolor}
\usepackage[hidelinks]{hyperref}

\begin{document}

\title{Emergence of Gravity's Dynamical and Topological Sectors from Pre-geometry}

\author{Andrea Addazi}
	\email{addazi@scu.edu.cn}
	\affiliation{School of Physics and Astronomy, Anqing Normal University, Anqing 246011, People's Republic of China}
	\affiliation{Institute of Astronomy and Astrophysics, Anqing Normal University, Anqing 246133, People's Republic of China}
	\affiliation{Laboratori Nazionali di Frascati INFN, Frascati (Roma), Italy, EU}

\author{Giuseppe Meluccio}
    \email{giuseppe.meluccio-ssm@unina.it}
    \affiliation{Scuola Superiore Meridionale, Largo San Marcellino 10, Napoli 80138, Italy}
    \affiliation{INFN Sezione di Napoli, Complesso Universitario di Monte Sant'Angelo, Edificio 6, Via Cintia 21, Napoli 80126, Italy}
    \affiliation{Department of Physics, Loyola Marymount University, 1 LMU Drive, Los Angeles 90045, California, USA}

\date{\today}

\begin{abstract}
    We identify the complete set of fundamental building blocks for a 4D pre-geometric theory of gravity. Based on a gauge theory of $SO(1,4)$ or $SO(2,3)$ coupled to a Higgs-like field under the rigid constraint of general covariance in the unbroken phase, these building blocks -- $\mathcal{L}_\textup{MM}$, $\mathcal{L}_\textup{W}$, $\mathcal{L}_{J}$, $\mathcal{L}_\theta$ and $\mathcal{L}_\vartheta$ -- constitute the minimal generating set of independent field monomials from which any pre-geometric action, including arbitrary functionals thereof, can be constructed. While the most general pre-geometric theory can extend beyond linear combinations, these five irreducible invariants serve as the `atomic' constituents of all possible pre-geometric dynamics. Upon spontaneous symmetry breaking, they collectively generate an emergent gravitational theory consisting of the Einstein--Hilbert action, the cosmological constant term and all 4D topological invariants: the Gauss--Bonnet, Pontryagin, Holst and Nieh--Yan terms. The unification of gravity's dynamical and topological sectors from a common pre-geometric source represents the central result of this work. We also uncover a see-saw mechanism linking the Planck mass and the cosmological constant, as well as several novel relations for the coupling constants of the topological sector, inclusive of the Barbero--Immirzi parameter. This framework establishes the pre-geometric foundations from which all aspects of gravitation can dynamically emerge, providing a unified starting point for quantum gravity, dark energy phenomenology and the study of topological phases in gravitational theories.
\end{abstract}

\maketitle

\section{Introduction}
The search for a quantum theory of gravity has long motivated the idea that Einstein's General Relativity (GR) might not be fundamental, but rather emerge from a more primitive structure. Among the most elegant realisations of this idea are gauge-theoretical approaches, wherein gravity emerges from the spontaneous symmetry breaking (SSB) of a gauge group larger than the Lorentz one $SO(1,3)$. Seminal contributions in this direction include the MacDowell--Mansouri (MM) \cite{macdowell:unified,SA1,SA2} and Wilczek (W) \cite{wilczek:gauge} formulations, which demonstrated how the Einstein--Hilbert (EH) action and the cosmological constant term can arise from a spontaneously broken gauge theory of the de Sitter group $SO(1,4)$ or the anti-de Sitter group $SO(2,3)$.\footnote{For additional contributions on related subjects, see for example Refs.\ \cite{Hehl:1995,Adler:1982,Krasnov:2012,Chamseddine:2016,Westman:2013,Westman:2014,Westman:2015,Zlosnik:2018,Koivisto:2019,gallagher:2022a,Koivisto:2023b,salgado:gauge,chamseddine:ordered}. In particular, when it comes to the choice of either $SO(1,4)$ or $SO(2,3)$, Ref.\ \cite{chamseddine:ordered} provided an argument based on the problem of coupling fermions to an (anti-)de Sitter tangent space.}

Inspired by this line of research, a key insight from the pre-geometric framework developed in our previous works \cite{addazi:pre-geometry,addazi:hamiltonian,meluccio:pregeometric,addazi:topological,addazi:holographic,capozziello:origin} is that both energy scales of gravity -- the Planck mass and the vacuum energy -- are emergent. The main formal feature of Pre-geometric Gravity is the absence of a fundamental spacetime metric. Before the SSB, the theory is in a sense scale-invariant, devoid of any standard notions of length or energy. The symmetry-breaking scale, set by the vacuum expectation value (VEV) of a Higgs-like field in the fundamental representation of $SO(1,4)$ or $SO(2,3)$, then generates all dimensionful parameters of low-energy gravity, in addition to all geometric quantities of interest such as the metric of spacetime (see Ref.\ \cite{addazi:pre-geometry}). The pre-geometric gauge potential gives rise to all gravitational degrees of freedom in the spontaneously broken phase, thus yielding a completely emergent Einstein--Cartan theory.

In this paper, we perform a systematic identification of the \emph{pre-geometric} fundamental building blocks -- the irreducible monomials in the pre-geometric fields and their first derivatives -- that are consistent with the rigidity of the principles of Pre-geometric Gravity (including that of general covariance in the unbroken phase), as articulated in Ref.\ \cite{addazi:pre-geometry}. A crucial clarification is in order regarding what we mean by ``fundamental building blocks''. The most general 4D pre-geometric action is not merely a linear combination of the terms we identify; rather, it is an arbitrary functional of these basic invariants. However, any such functionals can be expanded or constructed from a finite set of \emph{irreducible} generating monomials. Thus, the five Lagrangian densities we hereby present -- $\mathcal{L}_\textup{MM}$, $\mathcal{L}_\textup{W}$, $\mathcal{L}_J$, $\mathcal{L}_\theta$, and $\mathcal{L}_\vartheta$ -- form the minimal set of independent field monomials that serve as the `atomic' constituents of any 4D pre-geometric action. While one may consider functionals $f[\mathcal{L}_\textup{MM},\mathcal{L}_\textup{W},\mathcal{L}_J,\mathcal{L}_\theta,\mathcal{L}_\vartheta]$, these inevitably reduce -- upon SSB or in a derivative expansion -- to combinations of the effective couplings generated by the five fundamental building blocks \cite{capozziello:origin}. Therefore, identifying these building blocks is the first and essential step towards constructing and classifying the most general pre-geometric theory, including its nonlinear or non-perturbative extensions.

The pre-geometric framework can naturally accommodate the emergence of topological terms too. Indeed, the principal result of this work is the demonstration that the five pre-geometric building blocks collectively unify gravity's dynamical and topological sectors. Specifically, their SSB generates an emergent gravitational theory that contains not only the Einstein--Hilbert action with a cosmological constant term but also the complete set of 4D topological invariants: the Gauss--Bonnet (GB), Pontryagin (P), Holst (H) and Nieh--Yan (NY) terms. This unification from a common pre-geometric source is, to our knowledge, unprecedented in the literature.

Within this unified framework, we also uncover several novel results for the emergent coupling constants of the gravitational interaction. These relations unveil deep connections between the dynamical and the topological sectors of gravity, which in a special case are inclusive of a see-saw mechanism between the Planck mass and the cosmological constant; the Gauss--Bonnet coupling being proportional to the entropy of a de Sitter universe; the Pontryagin and Nieh--Yan couplings being proportional, with the former being in turn inversely proportional to the cosmological constant and the latter being in turn directly proportional to the cosmological constant; the Barbero--Immirzi parameter being proportional to the Gauss--Bonnet coupling and independent of the VEV scale.

This article is organised as follows. In Sec.\ \ref{sec:2} we introduce the notation and present the five irreducible Lagrangian densities that form the generating set of any 4D pre-geometric action. We compute the SSB for each term, derive the emergent gravitational theory, and analyse the relations between pre-geometric and emergent coupling constants. In Sec.\ \ref{sec:3} we then discuss a special case that highlights the intrinsic relations (such as the see-saw mechanisms) underpinning the emergent gravitational parameters. Sec.\ \ref{sec:4} concludes the paper by summarising our main findings, discussing the implications for dynamical dark energy and quantum gravity via functional extensions that promote couplings to dynamical gravi-axion fields, and outlining directions for future research, including the quest for a UV conformal fixed point in the pre-geometric theory \cite{addazi:topological,addazi:prospective,addazi:solution, addazi:holographic}.

\section{The Most General 4D Topological Extension of Einstein Gravity}\label{sec:2}
The pre-geometry of spacetime is determined by the coupled dynamics of the gauge potential $A$ of the theory and a Higgs-like field $\phi$; the former is the gauge field of either $SO(1,4)$ or $SO(2,3)$, while the latter is a scalar field in the fundamental representation of the chosen gauge group. Whenever a double sign is encountered in the text, the first sign refers to the choice of $SO(1,4)$ and the second one to the choice of $SO(2,3)$. In this work we employ natural units and use the same notation of Ref.\ \cite{meluccio:pregeometric}: uppercase Latin letters $A,B,C$ etc.\ run from $0$ to $4$ and lowercase Latin letters $a,b,c$ etc.\ run from $0$ to $3$, representing tangent space (or internal) indices in the unbroken and the spontaneously broken phases respectively; Greek letters $\lambda,\mu,\nu$ etc.\ run from $0$ to $3$, indicating instead spacetime (or external) indices. The internal space metric $\eta$ has signature $(-,+,+,+,\pm)$ before the SSB and $(-,+,+,+)$ after the SSB. The Levi-Civita symbol and tensor are denoted respectively by $\epsilon$ and $\varepsilon$, and their relation is $\varepsilon^{\mu\nu\rho\sigma}=\epsilon^{\mu\nu\rho\sigma}/\sqrt{-g}$ with $g$ being the metric determinant ($\varepsilon_{ABCDE}=\epsilon_{ABCDE}$ on tangent spaces).

When the Higgs-like field acquires a nonzero VEV
\begin{equation}\label{eq:SSB_phi}
    \phi^A\xrightarrow{SSB}v\delta_4^A
\end{equation}
as a result of a dynamical mechanism of SSB (see Ref.\ \cite{addazi:pre-geometry} for details about the potential), the gauge connection decomposes into its spin and tetrad parts:
\begin{equation}\label{eq:SSB_A}
    A_\mu^{ab}\xrightarrow{SSB}\omega_\mu^{ab},\qquad A_\mu^{a4}\xrightarrow{SSB}me_\mu^a,
\end{equation}
where $m$ is a mass parameter introduced to make the tetrad fields dimensionless. In turn, the decomposition of the curvature tensor $F=dA+A\wedge A$ induced by the SSB is
\begin{equation}
    F_{\mu\nu}^{ab}\xrightarrow{SSB}R_{\mu\nu}^{ab}\mp2m^2e_{[\mu}^ae_{\nu]}^b,\qquad F_{\mu\nu}^{a4}\xrightarrow{SSB}mT_{\mu\nu}^a,
\end{equation}
where $R$ and $T$ are the Riemann and torsion tensors respectively. We employ the curved-spacetime definition of the Hodge star operator, so for example the Hodge dual of a two-form $F$ is given by
\begin{multline}
    F^{ab}=\frac{1}{2}F_{\mu\nu}^{ab}dx^\mu\wedge dx^\nu,\\
    \Tilde{F}^{ab}\equiv(\star F)^{ab}=\frac{1}{2}(\star F)_{\mu\nu}^{ab}dx^\mu\wedge dx^\nu,
\end{multline}
where
\begin{equation}
    \Tilde{F}_{\mu\nu}^{ab}\equiv(\star F)_{\mu\nu}^{ab}=\frac{\sqrt{-g}}{2}g^{\rho\tau}g^{\sigma\xi}\epsilon_{\mu\nu\tau\xi}F_{\rho\sigma}^{ab}.
\end{equation}
For later convenience, we define the following geometric scalars:
\begin{subequations}
    \begin{align}
        \mathcal{G}&\equiv R^2-4R_{\mu\nu}R^{\mu\nu}+R_{\mu\nu\rho\sigma}R^{\mu\nu\rho\sigma}\nonumber\\
        &=-\frac{1}{4}\varepsilon_{abcd}\varepsilon^{\mu\nu\rho\sigma}R_{\mu\nu}^{ab}R_{\rho\sigma}^{cd}=-\frac{1}{2}\varepsilon_{\mu\nu\rho\sigma}R^{\mu\nu}_{\phantom{\tau\xi}\tau\xi}\Tilde{R}^{\rho\sigma\tau\xi},\\
        \mathcal{P}&\equiv\frac{1}{2}\varepsilon^{\mu\nu\rho\sigma}R^\tau_{\phantom{\tau}\xi\mu\nu}R^\xi_{\phantom{\xi}\tau\rho\sigma}=-R_{\mu\nu\rho\sigma}\Tilde{R}^{\mu\nu\rho\sigma},\\
        \Tilde{R}&\equiv\frac{1}{2}\varepsilon^{\mu\nu\rho\sigma}R_{\mu\nu\rho\sigma}=e^\mu_ae^\nu_b\Tilde{R}_{\mu\nu}^{ab}=\Tilde{R}^{\mu\nu}_{\phantom{\mu\nu}\mu\nu},\\
        \mathcal{T}&\equiv\frac{1}{2}\varepsilon^{\mu\nu\rho\sigma}T^\lambda_{\phantom{\lambda}\mu\nu}T_{\lambda\rho\sigma}=T_{\lambda\mu\nu}\Tilde{T}^{\lambda\mu\nu},\\
        \mathcal{N}&\equiv\Tilde{R}-\frac{1}{2}\mathcal{T}.\label{eq:N}
    \end{align}
\end{subequations}
Eq.\ \eqref{eq:SSB_phi} neglects any excitations of the Higgs-like field $\phi$, which come in the form of a scalar degree of freedom $\rho$ to be added to the VEV $v$ in the unitary gauge. Such scalar field $\rho$ is expected to have a Planckian mass in order for the phase transition from a pre-geometric to a metric spacetime to happen at the alleged scale of quantum gravity \cite{addazi:pre-geometry}. Eq.\ \eqref{eq:SSB_phi} is thus formally exact only in regimes well below the Planck energy, in which $\rho$ is effectively frozen out.

By requiring the principle of general covariance to hold in the unbroken phase too (see Ref.\ \cite{addazi:pre-geometry} for a more in-depth discussion), we can write down \emph{only} five different pre-geometric Lagrangian densities involving the pre-geometric fields $A$ and $\phi$ and their first derivatives (in a polynomial way):
\begin{subequations}\label{eq:L_PG}
    \begin{align}
        \mathcal{L}_\textup{MM}&\equiv k_\textup{MM}\epsilon_{ABCDE}\epsilon^{\mu\nu\rho\sigma}F_{\mu\nu}^{AB}F_{\rho\sigma}^{CD}\phi^E,\\
        \mathcal{L}_\textup{W}&\equiv k_\textup{W}\epsilon_{ABCDE}\epsilon^{\mu\nu\rho\sigma}F_{\mu\nu}^{AB}\nabla_\rho\phi^C\nabla_\sigma\phi^D\phi^E,\\
        \mathcal{L}_J&\equiv k_J\epsilon_{ABCDE}\epsilon^{\mu\nu\rho\sigma}\nabla_\mu\phi^A\nabla_\nu\phi^B\nabla_\rho\phi^C\nabla_\sigma\phi^D\phi^E,\\
        \mathcal{L}_\theta&\equiv k_\theta\eta_{AC}\eta_{BD}\epsilon^{\mu\nu\rho\sigma}F_{\mu\nu}^{AB}F_{\rho\sigma}^{CD},\\
        \mathcal{L}_\vartheta&\equiv k_\vartheta\eta_{AC}\eta_{BD}\epsilon^{\mu\nu\rho\sigma}F_{\mu\nu}^{AB}\nabla_\rho\phi^C\nabla_\sigma\phi^D.
    \end{align}
\end{subequations}
Through Eqs.\ \eqref{eq:SSB_phi} and \eqref{eq:SSB_A}, their SSB is given respectively by
\begin{subequations}
    \begin{align}
        \mathcal{L}_\textup{MM}&\xrightarrow{SSB}4k_\textup{MM}v(\pm4m^2R-24m^4-\mathcal{G})\sqrt{-g},\\
        \mathcal{L}_\textup{W}&\xrightarrow{SSB}4k_\textup{W}v^3(-m^2R\pm12m^4)\sqrt{-g},\\
        \mathcal{L}_J&\xrightarrow{SSB}-24k_Jv^5m^4\sqrt{-g},\\
        \mathcal{L}_\theta&\xrightarrow{SSB}-2k_\theta(\mathcal{P}\pm4m^2\Tilde{R}\mp2m^2\mathcal{T})\sqrt{-g},\label{eq:L_theta-SSB}\\
        \mathcal{L}_\vartheta&\xrightarrow{SSB}2k_\vartheta v^2m^2\Tilde{R}\sqrt{-g}.
    \end{align}
\end{subequations}
The mass dimensions of the five pre-geometric coupling constants are $[k_\textup{MM}]=[M]^{-1}$, $[k_\textup{W}]=[M]^{-3}$, $[k_J]=[M]^{-5}$, $[k_\theta]=[M]^0$, $[k_\vartheta]=[M]^{-2}$. Notice that, in the absence of a fundamental spacetime metric, the external indices of all pre-geometric terms in Eq.\ \eqref{eq:L_PG} are contracted by means of the contravariant Levi-Civita symbol.

The complete pre-geometric Lagrangian density is thus
\begin{equation}
    \mathcal{L}_\textup{PG}=\mathcal{L}_\textup{MM}+\mathcal{L}_\textup{W}+\mathcal{L}_J+\mathcal{L}_\theta+\mathcal{L}_\vartheta.
\end{equation}
Its SSB yields the gravitational Lagrangian
\begin{equation}\label{eq:L_SSB1}
    \begin{split}
        \frac{\mathcal{L}_\textup{G}}{\sqrt{-g}}&=\frac{M_\textup{P}^2}{2}R-M_\textup{P}^2\Lambda\\
        &+\theta_\textup{GB}\mathcal{G}+\theta_\textup{P}\mathcal{P}+\frac{M_\textup{P}^2}{2\gamma'}\Tilde{R}+\theta_T\mathcal{T},
    \end{split}
\end{equation}
once the six emergent coupling constants ($M_\textup{P}$, $\Lambda$, $\theta_\textup{GB}$, $\theta_\textup{P}$, $\gamma'$ and $\theta_T$) of the gravitational interaction are identified as follows in terms of the seven pre-geometric constants ($k_\textup{MM}$, $k_\textup{W}$, $k_J$, $k_\theta$, $k_\vartheta$, $v$ and $m$) that characterise $\mathcal{L}_\textup{PG}$:
\begin{subequations}\label{eq:constants1}
    \begin{align}
        M_\textup{P}^2&\equiv8vm^2(\pm4k_\textup{MM}-k_\textup{W}v^2),\\
        \Lambda&\equiv3m^2\frac{4k_\textup{MM}\mp2k_\textup{W}v^2+k_Jv^4}{\pm4k_\textup{MM}-k_\textup{W}v^2},\\
        \theta_\textup{GB}&\equiv-4k_\textup{MM}v,\\
        \theta_\textup{P}&\equiv-2k_\theta,\\
        \gamma'&\equiv\frac{2v(\pm4k_\textup{MM}-k_\textup{W}v^2)}{\mp4k_\theta+k_\vartheta v^2},\\
        \theta_T&\equiv\pm4k_\theta m^2.
    \end{align}
\end{subequations}
Newton's gravitational constant is defined as $G=1/(8\pi M_\textup{P}^2)$. Different observations can be made on these results.
\begin{enumerate}
    \item The (reduced) Planck mass $M_\textup{P}$ receives contributions only from $\mathcal{L}_\textup{MM}$ and $\mathcal{L}_\textup{W}$, which implies that at least one of the two is indispensable for the emergent gravitational theory to be realistic.
    \item The cosmological constant $\Lambda$ receives contributions only from $\mathcal{L}_\textup{MM}$, $\mathcal{L}_\textup{W}$ and $\mathcal{L}_J$, with the latter contributing exclusively to $\Lambda$.
    \item $\mathcal{L}_\theta$ and $\mathcal{L_\vartheta}$ do not contribute to either $M_\textup{P}$ or $\Lambda$.
    \item $\theta_\textup{GB}$ receives contributions only from $\mathcal{L}_\textup{MM}$.
    \item $\theta_\textup{P}$ receives contributions only from $\mathcal{L}_\theta$.
    \item The contribution of the Barbero--Immirzi parameter $\gamma'$ depends crucially on the difference between $k_\vartheta v^2$ and $\pm4k_\theta$, and thus on the contributions from $\mathcal{L}_\theta$ and $\mathcal{L_\vartheta}$, though it is related to $\mathcal{L}_\textup{MM}$ and $\mathcal{L}_\textup{W}$ too via the definition of $M_\textup{P}$.
    \item $\theta_T$ receives contributions only from $\mathcal{L}_\theta$ and determines the magnitude of the only term in the emergent gravitational theory which depends explicitly on torsion.
    \item The ratio between the terms proportional to $\Tilde{R}$ and to $\mathcal{T}$ emerging from $\mathcal{L}_\theta$ (see Eq.\ \eqref{eq:L_theta-SSB}) is the same as that between the same quantities in the definition of the scalar $\mathcal{N}$ (in Eq.\ \eqref{eq:N}), meaning that the contribution of $\mathcal{L}_\theta$ can always be expressed in terms of $\mathcal{P}$ and $\mathcal{N}$ rather than $\mathcal{P}$, $\Tilde{R}$ and $\mathcal{T}$.
\end{enumerate}

The last two observations reveal that the last two terms of the emergent theory \eqref{eq:L_SSB1}, which are proportional to $\Tilde{R}$ and $\mathcal{T}$ respectively, can be recast in an equivalent way as two different terms, proportional to $\Tilde{R}$ and $\mathcal{N}$, by simply redefining their emergent coupling constants $\gamma'$ and $\theta_T$ as some new constants $\gamma$ and $\theta_\textup{NY}$. In other words, the SSB of the complete pre-geometric Lagrangian density can also be expressed as
\begin{equation}\label{eq:L_SSB2}
    \begin{split}
        \frac{\mathcal{L}_\textup{G}}{\sqrt{-g}}&=\frac{M_\textup{P}^2}{2}R-M_\textup{P}^2\Lambda\\
        &+\theta_\textup{GB}\mathcal{G}+\theta_\textup{P}\mathcal{P}+\frac{M_\textup{P}^2}{2\gamma}\Tilde{R}+\theta_\textup{NY}\mathcal{N},
    \end{split}
\end{equation}
where this time
\begin{subequations}\label{eq:constants2}
    \begin{align}
        M_\textup{P}^2&\equiv8vm^2(\pm4k_\textup{MM}-k_\textup{W}v^2),\\
        \Lambda&\equiv3m^2\frac{4k_\textup{MM}\mp2k_\textup{W}v^2+k_Jv^4}{\pm4k_\textup{MM}-k_\textup{W}v^2},\\
        \theta_\textup{GB}&\equiv-4k_\textup{MM}v,\\
        \theta_\textup{P}&\equiv-2k_\theta,\\
        \gamma&\equiv\frac{2(\pm4k_\textup{MM}-k_\textup{W}v^2)}{k_\vartheta v},\\
        \theta_\textup{NY}&\equiv\mp8k_\theta m^2.
    \end{align}
\end{subequations}
Comparing Eqs.\ \eqref{eq:constants1} and \eqref{eq:constants2} shows that the effect of the above redefinition of parameters is to disentangle the contributions of $\mathcal{L}_\theta$ and $\mathcal{L_\vartheta}$: now the latter contributes only to $\gamma$ and the former contributes only to $\theta_\textup{P}$ and $\theta_\textup{NY}$. The expression \eqref{eq:L_SSB2} also highlights the main result of this paper: the gravitational theory $\mathcal{L}_\textup{G}$ that emerges from the pre-geometric theory $\mathcal{L}_\textup{PG}$ is the most general 4D \emph{topological} extension of Einstein gravity. Indeed, its first two terms describe the Einstein--Cartan theory as a sum of the Einstein--Hilbert and the cosmological constant terms respectively, while the other four coincide with \emph{all} possible topological terms that can be defined on a 4D metric spacetime, i.e.\ the Gauss--Bonnet, Pontryagin, Holst and Nieh--Yan terms respectively.\footnote{This list of topological terms excludes boundary terms, such as the Gibbons--Hawking--York boundary term (which is ostensibly hard to express in a pre-geometric form, if that is possible at all).} In formulae,
\begin{widetext}
    \begin{equation}\label{eq:total-pre-geometric-action}
        S_\textup{PG}=\int\mathcal{L}_\textup{PG}\,d^4x=S_\textup{MM}+S_\textup{W}+S_J+S_\theta+S_\vartheta\xrightarrow{SSB}S_\textup{G}=\int\mathcal{L}_\textup{G}\sqrt{-g}\,d^4x=S_\textup{EH}+S_\Lambda+S_\textup{GB}+S_\textup{P}+S_\textup{H}+S_\textup{NY},
    \end{equation}
\end{widetext}
where the pre-geometric actions are given (together with Eqs.\ \eqref{eq:L_PG}) by
\begin{subequations}
    \begin{align}
        S_\textup{MM}&=\int\mathcal{L}_\textup{MM}\,d^4x\nonumber\\
        &=-4k_\textup{MM}\int\epsilon_{ABCDE}F^{AB}\wedge F^{CD}\phi^E,\\
        S_\textup{W}&=\int\mathcal{L}_\textup{W}\,d^4x\nonumber\\
        &=-2k_\textup{W}\int\epsilon_{ABCDE}F^{AB}\wedge\nabla\phi^C\wedge\nabla\phi^D\phi^E,\\
        S_J&=\int\mathcal{L}_J\,d^4x\nonumber\\
        &=-k_J\int\epsilon_{ABCDE}\nabla\phi^A\wedge\nabla\phi^B\wedge\nabla\phi^C\wedge\nabla\phi^D\phi^E,\\
        S_\theta&=\int\mathcal{L}_\theta\,d^4x=-4k_\theta\int F_{AB}\wedge F^{AB},\\
        S_\vartheta&=\int\mathcal{L}_\vartheta\,d^4x=-2k_\vartheta\int F_{AB}\wedge\nabla\phi^A\wedge\nabla\phi^B,
    \end{align}
\end{subequations}
while the emergent gravitational actions are given (together with Eqs.\ \eqref{eq:constants2}) by
\begin{subequations}
    \begin{align}
        S_\textup{EH}&=\frac{M_\textup{P}^2}{2}\int R\sqrt{-g}\,d^4x=\frac{M_\textup{P}^2}{2}\int\star(e_a\wedge e_b)\wedge R^{ab},\\
        S_\Lambda&=-M_\textup{P}^2\Lambda\int\sqrt{-g}\,d^4x=\frac{M_\textup{P}^2\Lambda}{4}\int\star e_a\wedge e^a,\\
        S_\textup{GB}&=\theta_\textup{GB}\int\mathcal{G}\sqrt{-g}\,d^4x=\theta_\textup{GB}\int\epsilon_{abcd}R^{ab}\wedge R^{cd},\\
        S_\textup{P}&=\theta_\textup{P}\int\mathcal{P}\sqrt{-g}\,d^4x=2\theta_\textup{P}\int R_{ab}\wedge R^{ab},\\
        S_\textup{H}&=\frac{M_\textup{P}^2}{2\gamma}\int\Tilde{R}\sqrt{-g}\,d^4x=-\frac{M_\textup{P}^2}{2\gamma}\int e_a\wedge e_b\wedge R^{ab},\\
        S_\textup{NY}&=\theta_\textup{NY}\int\mathcal{N}\sqrt{-g}\,d^4x\nonumber\\
        &=\theta_\textup{NY}\int(T_a\wedge T^a-e_a\wedge e_b\wedge R^{ab}).
    \end{align}
\end{subequations}

A clarifying remark on the topological nature of the Holst term is due here. The Gauss--Bonnet, Pontryagin and Nieh--Yan \cite{nieh:NY} terms are topological because they can be recast as total derivatives, given that respectively
\begin{subequations}
    \begin{align}
        \epsilon_{abcd}R^{ab}\wedge R^{cd}&=d\biggl[\epsilon_{abcd}\omega^{ab}\wedge\biggl(d\omega^{cd}+\frac{2}{3}\omega^c_{\phantom{c}e}\wedge\omega^{ed}\biggr)\biggr],\\
        R_{ab}\wedge R^{ab}&=d\biggl[\omega_{ab}\wedge\biggl(d\omega^{ab}-\frac{2}{3}\omega^b_{\phantom{b}c}\wedge\omega^{ca}\biggr)\biggr],\\
        T_a\wedge T^a&-e_a\wedge e_b\wedge R^{ab}=d(e^a\wedge T_a).
    \end{align}
\end{subequations}
Instead, the Holst term is topological only in the absence of torsion. This is easily understood from Eq.\ \eqref{eq:N}, whereby $\Tilde{R}=\mathcal{N}$ when $\mathcal{T}=0$. That said, in a pre-geometric theory of gravity there is no a priori reason why torsion should vanish. Therefore, in general the emergent gravitational theory under exam will represent a topological extension of the Einstein--Cartan theory only if $k_\vartheta=0$, i.e.\ if the pre-geometric term $\mathcal{L}_\vartheta$ is not included.

\section{A Special Case:\linebreak The Link Between Dynamical and Topological Sectors of Gravity}\label{sec:3}
We can draw more insight from the SSB of the pre-geometric theory \eqref{eq:total-pre-geometric-action} by studying a special case, in which only two of the pre-geometric coupling constants are independent:
\begin{equation}\label{eq:special-case}
    k_\textup{PG}\equiv\pm k_\textup{MM}=k_\textup{W}v^2=\pm k_Jv^4,\qquad k_\Theta\equiv\pm k_\theta=k_\vartheta v^2.
\end{equation}
This is interesting because it allows to group the pre-geometric terms based on whether the internal indices are contracted with the Levi-Civita symbol or the generalised Minkowski metric:
\begin{equation}
    \begin{split}
        \mathcal{L}_\textup{PG}&=k_\textup{PG}\epsilon_{ABCDE}\epsilon^{\mu\nu\rho\sigma}(\pm F_{\mu\nu}^{AB}F_{\rho\sigma}^{CD}+v^{-2}F_{\mu\nu}^{AB}\nabla_\rho\phi^C\nabla_\sigma\phi^D\\
        &\pm v^{-4}\nabla_\mu\phi^A\nabla_\nu\phi^B\nabla_\rho\phi^C\nabla_\sigma\phi^D)\phi^E\\
        &+k_\Theta\eta_{AC}\eta_{BD}\epsilon^{\mu\nu\rho\sigma}F_{\mu\nu}^{AB}(\pm F_{\rho\sigma}^{CD}+v^{-2}\nabla_\rho\phi^C\nabla_\sigma\phi^D).
    \end{split}
\end{equation}
In this special case, the emergent gravitational coupling constants are
\begin{subequations}
    \begin{align}
        M_\textup{P}^2&\equiv24k_\textup{PG}vm^2,\\
        \Lambda&\equiv\pm3m^2,\\
        \theta_\textup{GB}&\equiv\mp4k_\textup{PG}v,\\
        \theta_\textup{P}&\equiv\mp2k_\Theta,\\
        \gamma&\equiv6\frac{k_\textup{PG}}{k_\Theta}v,\\
        \theta_\textup{NY}&\equiv-8k_\Theta m^2.
    \end{align}
\end{subequations}
Also this time, we can make several observations.
\begin{enumerate}
    \item The definition of the emergent Planck mass requires $k_\textup{PG}>0$.
    \item The cosmological constant is independent of all pre-geometric coupling constants and its sign is positive for $SO(1,4)$ or negative for $SO(2,3)$. There is a see-saw mechanism between the Planck mass and the cosmological constant (as already noted, for instance, in Refs.\ \cite{wilczek:gauge,addazi:pre-geometry}), since
    \begin{equation}
        \frac{M_\textup{P}^2}{\abs{\Lambda}}=8k_\textup{PG}v.
    \end{equation}
    \item The intensity of the Gauss--Bonnet term is set by the ratio between the Planck mass and the cosmological constant, as
    \begin{equation}\label{eq:GB}
        \theta_\textup{GB}=-\frac{M_\textup{P}^2}{2\Lambda}.
    \end{equation}
    \item The Pontryagin and Nieh--Yan coupling constants are proportional, with the proportionality constant set by the cosmological constant:
    \begin{equation}
        \frac{\theta_\textup{NY}}{\theta_\textup{P}}=\frac{4}{3}\Lambda.
    \end{equation}
    \item Just like $\theta_\textup{P}$ and $\theta_\textup{NY}$, also the Barbero--Immirzi parameter depends only on the free parameter $k_\Theta$, as it can be recast as
    \begin{equation}
        \gamma=\frac{3M_\textup{P}^2}{4k_\Theta\abs{\Lambda}}=\frac{3\theta_\textup{GB}}{\theta_\textup{P}}.
    \end{equation}
    Therefore, the three parameters $\theta_\textup{P}$, $\gamma$ and $\theta_\textup{NY}$ are either all real or all complex, depending on the imaginary part of $k_\Theta$ (while $M_\textup{P}$, $\Lambda$ and $\theta_\textup{GB}$ are all real). It is worth noticing that the dependence of $\gamma$ on the VEV scale $v$ is effectively cancelled by the knowledge of the other emergent parameters. Additionally, in this case $\gamma$ ends up being directly proportional to $\theta_\textup{GB}$ and inversely proportional to $\theta_\textup{P}$. Indeed, the above relation can also be inverted to give
    \begin{equation}
        \theta_\textup{P}=-\frac{3M_\textup{P}^2}{2\gamma\Lambda}.
    \end{equation}
    \item The Nieh--Yan coupling constant is proportional to the cosmological constant, as
    \begin{equation}
        \theta_\textup{NY}=-\frac{8}{3}k_\Theta\abs{\Lambda}.
    \end{equation}
\end{enumerate}

With regards to the third observation, the emergence of the Gauss--Bonnet term is an intriguing feature of the MM formulation, which sets it apart from the W one. As a matter of fact, the relation \eqref{eq:GB} for $\theta_\textup{GB}$ can have profound consequences for the vacuum structure of the emergent theory -- see Ref.\ \cite{addazi:solution} for a dedicated study. In essence, the Gauss--Bonnet coupling acts as a gravitational $\theta$-angle, endowing a suitable symmetry-breaking potential of the Higgs-like field with a large, discrete symmetry that forces its vacuum to be degenerate. Each degenerate vacuum state is specified by an integer $k$ and corresponds to a different pair of values for $\Lambda$ and $\theta_\textup{GB}$. This leads to a natural see-saw mechanism: a large Gauss--Bonnet coupling (i.e.\ a large-$k$ sector) implies a hierarchically small cosmological constant. Moreover, the model presented in \cite{addazi:solution} naturally predicts that the cosmological constant is quantised, and that the de Sitter entropy is discretised and proportional to the large-$k$ VEV of $\phi$ -- in compliance with the holographic principle.

Observe that the subcase of Eq.\ \eqref{eq:special-case} with $k_\vartheta=0$ is tantamount to setting $k_\textup{W}=k_J=0$ and redefining the only two remaining pre-geometric coupling constants, $k_\textup{MM}$ and $k_\theta$. In other words, the most general 4D topological extension of the Einstein--Cartan theory (even in the presence of torsion) can emerge in the pre-geometric framework even only from the minimalistic theory $S_\textup{MM}+S_\theta$.

\section{Conclusions and Perspectives}\label{sec:4}
In this paper we have identified the five irreducible pre-geometric building blocks -- $\mathcal{L}_\textup{MM}$, $\mathcal{L}_\textup{W}$, $\mathcal{L}_J$, $\mathcal{L}_\theta$ and $\mathcal{L}_\vartheta$ -- that constitute the minimal generating set for any 4D pre-geometric theory of gravity based on the gauge group $SO(1,4)$ or $SO(2,3)$, under the rigid constraint of general covariance in the unbroken phase. While the most general pre-geometric action can involve arbitrary functionals of these building blocks \cite{capozziello:origin}, their linear combination already yields, upon SSB, a gravitational theory that unifies all known sectors of gravity, i.e.\ the Einstein--Hilbert action, the cosmological constant term and the complete set of 4D topological invariants: the Gauss--Bonnet, Pontryagin, Holst and Nieh--Yan terms. This emergent theory of gravity is nothing else but the most general 4D topological extension of the Einstein--Cartan theory. The main result of this article is thus the unification of gravity's dynamical and topological sectors emerging from a common pre-geometric origin.

Our findings reveal numerous surprising relations between the emergent gravitational couplings that were previously unnoticed in the literature. The form of these relations simplifies considerably in the special case studied in Sec.\ \ref{sec:3}, which consequently allows for an easier physical understanding of the emergent theory. The first result is a see-saw mechanism between the Planck mass $M_\textup{P}$ and the cosmological constant $\Lambda$, demonstrating how a relatively large Planck scale and a relatively small cosmological constant can arise from the pre-geometric dynamics and, in particular, from the same symmetry-breaking scale $v$ set by the Higgs-like field $\phi$.

A second result concerns the Gauss--Bonnet invariant, which can originate solely from $\mathcal{L}_\textup{MM}$. Its emergent coupling $\theta_\textup{GB}$ is determined by the (relatively large) ratio between $M_\textup{P}^2$ and $\Lambda$ -- which is of the same order of magnitude as the fine-tuning required to fix the infamous cosmological constant problem \cite{weinberg:cosmological}. This also means that $\theta_\textup{GB}$ is proportional to the entropy of a de Sitter universe \cite{addazi:holographic,addazi:solution}.

Third, we have uncovered a striking and direct relation between the Pontryagin coupling $\theta_\textup{P}$ and the cosmological constant, as $\theta_\textup{P}\propto1/(\gamma\Lambda)$. This result mirrors the one anticipated from the Chern--Simons--Kodama state in Loop Quantum Gravity (LQG) \cite{Alexander:2025qkx}. Our relation introduces a correction with a factor of $\gamma$ in the law $\theta_\textup{P}\propto1/\Lambda$ predicted in Ref.\ \cite{Alexander:2025qkx}.

Speaking of the emergence of the Barbero--Immirzi parameter $\gamma$, we have shown that the Holst term, often considered to be an ambiguous addition to the gravitational action, arises naturally from the SSB of specific pre-geometric terms. The constant $\gamma$ is in principle a free parameter in the definition of the Ashtekar--Barbero connection of LQG \cite{Ashtekar:2021kfp}, and is only fixed \emph{a posteriori} to match the black hole entropy counting. A fourth result of our framework is instead that $\gamma$ is not arbitrary, rather it emerges as a precise combination of pre-geometric couplings; in particular, it is proportional to $\theta_\textup{GB}$. Crucially, we have proved that $\gamma$ can be independent of the VEV $v$, in sharp contrast with standard $\theta$-angles (which scale with $v$). This provides a potential resolution to the classical ambiguity of the Holst term -- which does not affect GR's field equations -- by granting it a well-defined pre-geometric origin, while still preserving its quantum effects that are vital to the LQG program. For example, our framework can explain why the Barbero--Immirzi parameter is real and close to unity when considered alongside a gravi-axion mechanism that simultaneously rotates the Gauss--Bonnet and Pontryagin $\theta$-angles too.

The Pontryagin and Nieh--Yan terms can only arise from $\mathcal{L}_\theta$. A fifth result comes in the form of the Nieh--Yan coupling $\theta_\textup{NY}$ being directly proportional to the cosmological constant. This pre-geometric prediction provides a new observational signature to investigate the topological structure of spacetime, potentially linking torsion phenomenology to the dark energy sector.

The implications of this work extend well beyond a classical picture of emergent gravity. Our framework suggests that what we believe to be the fundamental constants of gravitation -- Newton's gravitational constant, the cosmological constant, the Barbero--Immirzi parameter and the topological couplings -- are not independent but can all be derived from a set of pre-geometric parameters, which includes a single VEV.

A particularly exciting avenue for future research is to promote these pre-geometric coupling constants to dynamical degrees of freedom. For instance, considering functionals of the form $\mathcal{L}_Jf[\mathcal{L}_{\textup{MM},\textup{W},\theta,\vartheta}\mathcal{L}_J^{-1}]$ can turn these parameters into dynamical fields such as gravi-axions. As argued in our companion papers \cite{addazi:holographic,addazi:solution, addazi:prospective}, this possibility has profound implications for the cosmological constant problem and dynamical dark energy. For example, the $\theta$-angle nature of the Gauss--Bonnet coupling could lead to a relaxation mechanism for the vacuum energy, analogously to the axion solution to the strong CP problem -- but in this case applied to the gravitational sector. Other research areas that offer opportunities for phenomenological enquiries about the role of the Higgs-like field $\phi$ include inflation and dark matter.

Furthermore, the pre-geometric viewpoint naturally incorporates quantum gravity effects. The renormalisation group flow of the emergent couplings -- $G$, $\Lambda$, $\gamma$, $\theta_\textup{GB}$, etc.\ -- can be studied directly from the pre-geometric point of view. Recent analyses relying on the stochastic quantisation of pre-geometric gravity \cite{addazi:topological} strongly suggest the existence of a UV conformal fixed point. This indicates that Pre-geometric Gravity may be asymptotically safe, providing a well-defined UV completion of Einstein gravity. The discovery of such a fixed point, and the resulting renormalisation group trajectories, could lead to predictive relations among the emergent couplings at low energies, relations that in principle could be tested with cosmological observations.

Additionally, as observed in Sec.\ \ref{sec:2}, the present analysis neglected the role of the excitations $\rho$ of the Higgs-like field $\phi$ after the SSB. This means that our results are formally exact only at energy scales well below the Planckian regime, as $\rho$ is frozen out and (direct) quantum gravity effects are washed out. Such IR limit corresponds to the scenarios where gravitational experiments have been carried out with great precision, confirming Einstein's theory over and over again. This does not only justify our focus on the `classical' gravitational action $S_\textup{G}$ of Eq.\ \eqref{eq:total-pre-geometric-action} (where $\rho=0$), but also offers an illuminating perspective on the nature of topological invariants. Indeed, if the same logic is followed in reverse by passing from the IR limit of Einstein gravity to the UV regime of the Planck scale but just before the pre-geometric phase transition, then the well-known geometric theory of gravity will exhibit ultra-high-energy deviations due to the coupling of the excitation $\rho$ to the geometry of spacetime: such coupling (whose precise form was not presented here for conciseness) exists in both the dynamical and topological sectors, implying that also the terms that we are used to conceive as topologically invariant are actually dynamical whenever the field $\rho$ is excited near the Planck scale. In other words, the topological terms of the gravitational interaction are born dynamical from the pre-geometric phase transition, and only \emph{become} trivial at low enough energies far from the Planck scale. This profound insight sets the Pre-geometric Gravity paradigm apart from mere gauge-theoretical reformulations of gravity such as Gauge Gravitation Theory or MacDowell--Mansouri / (a)dS gravity, revealing that it predicts dynamical deviations from GR whose phenomenological consequences are yet to be adequately explored (for one such case study, see for instance Ref.\ \cite{addazi:DESI}).

A complete understanding of the quantum dynamics of Pre-geometric Gravity, including the full role of the topological terms and the fate of the Barbero--Immirzi parameter under renormalisation, requires a programmatic approach of research which extends beyond the purposes of this paper. However, the identification of the irreducible pre-geometric building blocks and the unified framework presented here lay the essential groundwork for all such future investigations. From these minimal pre-geometric seeds, all that is gravitational can emerge.

\acknowledgments
The Authors thank the anonymous Referees for several suggestions and remarks on these subjects. AA's work is supported by the Program for Innovative Research Team in Anqing Normal University. GM acknowledges the support of Istituto Nazionale di Fisica Nucleare (INFN), Sezione  di Napoli, \textit{Iniziativa Specifica} QGSKY. This publication is based upon work from COST Action CA21136 -- ``Addressing observational tensions in cosmology with systematics and fundamental physics (CosmoVerse)'', supported by COST (European Cooperation in Science and Technology).

\end{document}